\documentclass[12pt]{article}
\usepackage{graphicx}
\makeatletter
\long\def\@makecaption#1#2{%
  \vskip\abovecaptionskip   
  \sbox\@tempboxa{#1. #2}%
  \ifdim \wd\@tempboxa >\hsize
    #1. #2\par
  \else
    \global \@minipagefalse
    \hb@xt@\hsize{\box\@tempboxa\hfil}%
  \fi
  \vskip\belowcaptionskip}
\makeatother
 
\def\lvm{\leavevmode\hbox to\parindent{\hfill}}
\def\req#1{(\ref{#1})}

\def\d{\partial}   
\def\dT{\partial T}
\def\dt{\partial t}
\def\du{\partial u}
\def\dr{\partial r}
\def\dm{\partial m}

\def\dE{\partial E}

\def\dX{\partial X}

\def\L{\left} 
\def\R{\right}
\def\eps{\varepsilon}

\def\BE{\begin{equation}}
\def\EE{\end{equation}}
\def\BA{\begin{array}}
\def\EA{\end{array}}
\def\BAN{\begin{eqnarray}}
\def\EAN{\end{eqnarray}}

\def\ga{\mathrel{\mathpalette\fun >}}
\def\fun#1#2{\lower3.6pt\vbox{\baselineskip0pt\lineskip.9pt
\ialign{$\mathsurround=0pt#1\hfil##\hfil$\crcr#2\crcr\sim\crcr}}}
\def\msun{M_\odot}
\def\Msun{$\msun$}

\def\ltsima{$\; \buildrel < \over \sim \;$}
\def\ltsim{\lower.5ex\hbox{\ltsima}}
\def\gtsima{$\; \buildrel > \over \sim \;$}
\def\gtsim{\lower.5ex\hbox{\gtsima}}

\def\FIG #1 #2 [#3] #4\par{%
  \begin{figure}[!ht] \begin{center}%
    \vskip -3mm
    \includegraphics*[#3]{#2}%
    \\ 
    \caption{\label{#1}\small #4} 
    \vskip -3mm
  \end{center}\end{figure}%
}

\def\FIGG #1 #2 #3 [#4] #5\par{%
  \begin{figure}[!ht]
    \vskip -3mm
    \includegraphics*[#4]{#2}
    \hfill
    \includegraphics*[#4]{#3}
    \vskip -5mm
    \caption{\label{#1}\small #5}
  \end{figure} 
}

\def\FIGFOUR #1 #2 #3 #4 #5 [#6] #7\par{%
  \begin{figure}[!ht]
    \vskip -3mm
    \includegraphics*[#6]{#2}
    \hfill
    \includegraphics*[#6]{#3}
    \includegraphics*[#6]{#4}
    \hfill
    \includegraphics*[#6]{#5}
    \caption{\label{#1}\small #7}
  \end{figure} 
}

\title{Dynamics and Radiation of Young Type-Ia Supernova Remnants:  
Important Physical Processes}
 
\author{E.I.Sorokina${}^{1\star}$, S.I.Blinnikov${}^{1,2}$,
        D.I.Kosenko${}^{1,2}$, P.Lundqvist${}^3$\\}
\date{Astronomy Letters, vol.30, pp. 737-750.}
 
\begin{document}
\begin{titlepage}
\maketitle
\thispagestyle{empty}
\vskip 1cm
\centerline{${}^1$\it Sternberg Astronomical Institute, 
Universitetskii pr., 13, Moscow, 119992 Russia} 
\centerline{${}^2$\it Institute for Theoretical and Experimental
Physics, B.Cheremushkinskaya 25, Moscow, 117218  Russia}
\centerline{${}^3$\it Stockholm Observatory, Albanova,  Stockholm, Sweden}

\vskip 1cm
\centerline{Received by Astron.Lett on 25.04.2004}
\vskip 1cm

\centerline{Abstract}
\vskip 0.8cm
We examine and analyze the physical processes that should be taken
into account when modeling young
type-Ia supernova remnants (SNRs)  with ages of several
hundred years, in which there are forward  (propagating into an 
interstellar medium) and reverse (propagating into ejecta) shock waves. 
It is shown, that the energy losses in the metal-rich ejecta can be
essential for remnants already at this stage of evolution.
The influence of electron thermal conduction and the rate of the
energy exchange between electrons and ions on the temperature
distribution and the X-radiation from such remnants is studied.
The data for Tycho SNR from the XMM-Newton space X-ray telescope
have been employed for the comparison of calculations with observations.
\vskip 3cm
{\it $^\star$sorokina@sai.msu.su}
\end{titlepage}

\newpage

\section{Introduction} \lvm
\label{intr} Numerical simulations of supernova remnants (SNRs) have been
conducted already long ago. 
However, since the physics of these objects is very complex, 
it has not yet been completely included
in any computer program in the world. 
Moreover, different physical processes can be essential
at different stages of evolution.

In this paper, we will consider young SNRs that  have swept up 
interstellar gas with a mass of the order the ejecta  mass
during their expansion after the supernova explosion.
Two shock waves are formed  in the SNR at this evolutionary phase. 
One shock propagates outside, into the interstellar medium, 
thus increasing the mass of the swept up gas, 
while the other travels inward through the supernova ejecta
cooled because of the adiabatic expansion.
This reverse  reheats the ejecta to temperatures of $10^7$ - $10^9$~K.
Thus, an additional opportunity appears to study the matter ejected
during the supernova explosion, to investigate the distribution of density, 
of chemical elements and, possibly, to better understand 
the physics of explosion.

Self-similar solutions for this stage of evolution were proposed
more than two decades ago by Nadyozhin (1981, 1985) and Chevalier (1982).
These solutions do correctly predict the overall hydrodynamic structure of
a young SNR and formation of shocks in it. 
However, many physical processes, without which it is not possible to 
accurately predict peculiarities of SNR radiation, 
are not taken into account in those  solutions. 
For example, the self-similar solutions ignore the role of
unsteady ionization in SNR hydrodynamics.
The ionization depends on the entire previous evolution, and it is very essential for young SNRs, since it takes more than several hundred
years to establish ionization equilibrium there.
Therefore, numerical simulations with account of
nonequilibrium processes are necessary for the quantitative understanding 
of the details of SNR radiation.

The realization of this fact has led to the appearance of many works, 
aimed at studying the influence of various physical processes in SNRs 
(time-dependent ionization, possible difference in temperatures of electrons
and ions, the influence of radiative losses,
the account of electron thermal conduction and nonthermal
particles).
Here we make reference only to works on hydrodynamic modeling of the Tycho SNR.
Without claiming to present a complete list, we summarize these 
works in Table~\ref{Tychowork} by specifying which physical processes 
were taken into account in each of them.

The time-dependent ionization was considered in all cited papers,
but its coupling to hydrodynamics was treated quite differently.
E.g., Brinkmann et al. (1989) took this coupling into account quite
correctly, while Badenes et al. (2003) ignored it and employed
a rather crude approximation of hydrodynamics with constant adiabatic $\gamma$.

The detailed calculation of hydrodynamic evolution was not performed only by
Hamilton and Sarazin (1984b).
Instead, these authors took a self-similar solution from
their previous paper (Hamilton and Sarazin 1984a),
but, at the same time, they were the only authors before our work,
who took into account the radiative losses in young SNRs at the age of a few hundred years.

The following notation is used in the ``$T_e \; {\rm vs.}\; T_i$ ''
column: ``$1T$'' for the works in which the electron and
ion temperatures are assumed to be equal; ``$2T$''for the
works in which these temperatures are different, but
become equal in most cases only through Coulomb
collisions; and ``$3T$'' for the case where the electron gas
is divided into two components: a hot component collisionlessly
heated at the shock front by plasma instabilities and
a cold component emerging behind the shock front
due to particle collisions in plasma. Thermal conduction was
taken into account in a self-consistent way in none
of the previous works on Tycho SNR, but Itoh et al. (1988) proposed
a temperature equilibration algorithm that, to some
extent, simulates the action of thermal conduction.

In the present work, we include both the electron thermal conduction
and the radiative losses. Our results do show
that these processes can significantly change the hydrodynamic
structure of a SNR. However, since the ionization
is time dependent, they do not strongly affect the
radiation, because the ionization rate is approximately the same for all the actually possible diversity in the temperature distribution, and the ion composition does not reach equilibrium values in several hundred
years and, hence, is also similar for all temperatures.

Our calculations produce 
X-ray spectra and SNR surface brightness distributions
under different physical assumptions in different
models.
We compare these results with recent observations of the Tycho SNR,
for which data with high spatial and spectral resolutions
from the XMM-Newton space telescope are
available (Decourchelle et al. 2001).


\begin{table}[h!] 
\caption{Allowance for the physical processes in works on modeling the Tycho SNR.}
\label{Tychowork}
\begin{center}
\begin{tabular}{|l|c|c|c|c|c|}
\hline
Work    & Hydro- & Time-dependent  & Radiative  & Thermal  & $T_e$   \\
        & dynamics & ionization   & losses & conduction   &  vs. $T_i$ \\
\hline
Hamilton, Sarazin (1984b) & --     & +        & +       & --       & 2--3$T$ \\
\hline
Itoh et al. (1988)        & +      & +        & --      & -- (+)   & 2$T$ \\
\hline 
Brinkmann et al. (1989)   & +      & +        & --      & --       & 1$T$ \\
\hline 
Badenes et al. (2003)     & +      & +        & --      & --       & 2$T$ \\
\hline 
This paper                & +      & +        & +       & +        & 1--2$T$ \\
\hline
\end{tabular}
\end{center}
\end{table}

\section{Models of SNRs} \lvm

We constructed each SNR model from two parts:
the supernova ejecta and the surrounding interstellar
medium. Since our main goal here was to investigate
the effects of various physical processes on SNR
radiation, we took only two type-Ia supernova models
with markedly different properties of the ejecta:

\begin{itemize} 
\item the classical W7 model (Nomoto et al. 1984):
       $M_{WD}=1.38\msun$, 
       $E_0=1.2 \times10^{51}$~ergs, 
$M(^{56}\mbox{Ni}) = 0.6 \msun$;
\item one of the first three-dimensional models from MPA
(Reinecke et al. 2002), below we will denote it MR0: $M_{WD}=1.38 \msun$,  $E_0=4.6\times10^{50}$~ergs,
       $M(^{56}\mbox{Ni}) = 0.43 \msun$.
\end{itemize} 
Another essential difference between these models is
the different degree of gas mixing: the mixing is very strong in MR0, so that even the outermost layers of ejecta have up to 20\%  of iron, while in W7 iron almost disappears already in the layer with mass 
coordinate 1.1\Msun.
The simulations of light curves (Blinnikov, Sorokina 2004)
have shown that the combination of different explosion energy and effectiveness of
ejecta mixing is an essential factor, which yielded practically
identical velocity of photosphere, and, as consequence, similar light curves
for both models.
One should note, however, that in {\it UBV} filters  MR0 is in much better
agreement with
observations, while the decline
rate of the bolometric
light curve better agrees with W7. We assume that
a crucial factor for the radiation of an optically thin
SNR must be mainly the degree of mixing, which
differs significantly in the used models. The
reverse shock successively heats up ever deeper
layers of the ejecta, and the spectral evolution of
the SNR depends on the degree of mixing and the
abundances of various elements. In our subsequent
papers, we are planning to pay more attention to
the variety of models and to clarify how the model
properties can be reflected in the pattern of SNR
radiation.

We surrounded the ejecta by a gas at rest with cosmic composition
having constant temperature ($10^4$ K) and density. 
In order for the observed XMM-Newton X-ray flux from the
Tycho SNR (Decourchelle et al. 2001) at a distance of 2.3 kpc to be equal to
the computed flux, we assumed the ambient number density
to be $5 \cdot10^{-24}$/cm${}^3$. The computation began at the SNR age 
$\sim 10$ yr, when the density of the model outer ejecta mesh zones exceeded the
interstellar density by a factor of 10 to 100. We assumed that the ejecta ``did not feel'' the presence of interstellar matter and expanded
adiabatically before this time.

\section{Basic equations and physical processes} \nopagebreak
\subsection{The equations and the method} \lvm

The code {\sc supremna} was developed for the solution of problem.
We used this code to solve the following system of spherically
symmetric differential equations in the Lagrangean coordinates:
\BAN
 {\dr\over\dt}&=&u\,,\label{velo}\\
 {\dr\over\dm}&=&{1\over 4\pi r^2\rho}\,,\\
 {\du\over\dt}&=&\!
        {}-4\pi r^2\,{\d (P_e+P_i)\over\dm}-{Gm\over r^2}\,,
  \label{accel}\\
 \L({\dE_e\over\dT_e}\R)_\rho\,{\dT_e\over\dt}&=&\!
        {}-4\pi\,P_e\,{\d\over\dm}\L(r^2\,u\R)
          -\,4\pi\,{\d\over\dm}\L(r^2 F_{\rm cond}\R)\nonumber\\
      &&-\eps_r\,
        -\,{\d\eps_{ion}\over\dt}
        -\L({\dE_e\over\dX_e}\R) {\dX_e\over\dt}
        +\,{1\over\rho}\,\nu_{ie}\,
            k_b\L(T_i-T_e\R),
                                \label{tpe}\\
 \L({\dE_i\over\dT_i}\R)_\rho\,{\dT_i\over\dt}&=&\!
        {}-4\pi\,P_i\,{\d\over\dm}\L(r^2\,u\R)
          -\,{1\over\rho}\,\nu_{ie}\,
            k_b\L(T_i-T_e\R),
                                \label{tpi}\\
 {\d{\mathbf X}\over\dt} &=&
        {}f(T_e,\rho,{\mathbf X}).
             \label{dotX}
\EAN

Here $u$  is the velocity; $\rho$  is the density; $T_e$ and $T_i$ 
are the electron and ion temperatures; $P_e$ and $P_i$ are the respective 
pressures (taking into account artificial viscosity, see below section \ref{tie}); $E_e$ and $E_i$ are the thermal energies per unit mass of the gas element
at the Lagrangean coordinate $m$ (corresponding to mass within a radius $r$) at the moment of time $t$; 
$F_{\rm cond}$ is the energy flux due to the electron-electron and electron-ion thermal conduction; 
$\nu_{ie}$ is the electron-ion collision frequency per unit volume;
 $\eps_r$ is the radiation energy loss rate per unit mass of the gas element;
$\d\eps_{ion}/\dt$ accounts for the change in specific thermal energy of gas 
due to the change of ionization state;
${\mathbf X} = \{X_{\rm HI},X_{\rm HII},
X_{\rm HeI}, \dots,X_{\rm NiXXIX} \} $ is  the abundance vector  of all ions of all included elements relative to the total number of atoms and ions.
We introduce also $X_e=n_e/n_{b}$ for the number of electrons per baryon.
Below, we discuss the physical
processes described by these equations and show to what extent
their allowance is important when modeling the radiation
in young SNRs.

To solve the system of equations~\req{velo} -- \req{dotX}, we used
a new implicit finite-difference method that we developed. 
It is based on the method of lines for the equations of hydrodynamics,
when the spatial derivatives are approximated by finite differences,
and the emerging system of ordinary differential equations is solved
by Gear-type multistep predictor-corrector methods (Gear 1971; see also Arushanyan, Zaletkin 1990).
Previously, we used a similar method to compute the light curves of supernovae,
and its detailed description can be found in the paper
by Blinnikov et al.(1998). 
Our problem differs from that described by Blinnikov et al.(1998) by the
absence of multigroup transfer equations (we assume
the ejecta and the swept-up interstellar gas to be
transparent, so any emitted photon escapes from the system). 
The heat exchange proceeds in the
approximation of thermal conduction according to
Eq.\req{tpe}, which is also effectively solved by the method
of lines consistent with hydrodynamics. However,
compared to the work by Blinnikov et al.(1998),
the problem under consideration is significantly complicated
by the need for a kinetic calculation of the
ionization state for the gas in each zone.

The time it takes for ionization
to reach a steady state exceeds several hundred years after the passage of the shock front through a gas element under the conditions typical of
young SNRs. 
Therefore, the time-dependent ionization must be computed
to model the radiation of historical Tycho-like SNRs. 
Since the radiation in the metal-rich
ejecta can affect the dynamics even in young SNRs
(see below), it is inadequate first to perform a hydrodynamic
SNR calculation (as was done, e.g., by
Badenes et al. 2003), preserving the density and
temperature history, and only then to compute the
ionization state from these data. 
We had to include the ion kinetic calculation in the general scheme. 
The physical side of the kinetics of a moving multicharge
plasma is described, for example, in the book by Borovskij et al.(1995). 
The computational aspect of this problem is far from trivial. 
We solve the kinetic problem as follows by ensuring full consistency
of hydrodynamics with kinetics.

At each time step the ``hydrodynamic'' predictor approximates the
evolution of $T$ and $\rho$ from Eqs.~\req{velo} -- \req{tpi} in each Lagrangean zone 
by polynomial fits up to the 4th of order in time $t$.
Knowing ion abundances from the previous step, we obtain their new values at the following moment of time by solving the system of kinetic equations,
utilizing the polynomial fits for $T$ and $\rho$ specified
by the ``hydrodynamic'' predictor.
We solve the equations of ion kinetics \req{dotX} by another 
predictor-corrector algorithm (also the Gear's method), allowing it 
to perform the necessary number of steps in each mesh zone for
calculating the new values of abundances with the prescribed accuracy.
Next, we substitute them together with the predicted values of
hydrodynamic variables into the right-hand sides of the equations, with which the ``hydrodynamic'' corrector then works.
If the results of the hydrodynamic predictor and corrector diverge 
then the procedure is repeated with the reduced step or order of the method. 
In the case of successful convergence, the possibility of an automatic
increase in the step or order of the method is considered.

The calculations of kinetics were performed for all ions of 15 most
abundant elements (H, He, C, N, O, Ne, Na, Mg, Al, Si, S, Ar, Ca, Fe, Ni)
by taking into account the electron-impact
ionization, autoionization, photo- and dielectronic
recombination, and charge exchange (Arnaud and
Rothenflug 1985; Verner and Yakovlev 1990; Verner
and Ferland 1996; Seaton 1959; Shull and Steenberg
1982; Nussbaumer and Storey 1983).

\subsection{Electron thermal conduction} \lvm

We take the thermal conduction  flux  in Eq.~\req{tpe}  to be
\BE
F_{\rm cond} = - \; {k_e \nabla T \over 1 +|k_e \nabla T|/F_{\rm sat}}.
\label{fcond} 
\EE
Here we assume that the electron energy  $E_e$ cannot be transferred with
a speed higher than the sound speed in an ion or electron gas,  $c_{si}$
or $c_{se}$, depending on which of these speeds is lower (in most cases 
this is the ion sound speed; see, for example, 
Bobrova, Sasorov 1993).
In this way, we can indirectly take
into account the electrical neutrality of the plasma as
a whole (Imshennik and Bobrova 1997). 
Thus, the heat conduction flux with any temperature gradient cannot be higher
than its ``saturated'' value $F_{\rm sat}=E_e \min(c_{si},c_{se})$.

The coefficient of electron thermal conductivity from Eq.~\req{fcond} includes
both the electron-electron and electron-ion thermal conduction.
It is (Spitzer, H\"arm 1953) 
\BE
\label{condcoef} k_e={2 \, k^{7/2} \,T^{5/2} \, \xi(Z) \over e^4 \,m_e^{1/2} \,Z \ln \Lambda} \,
                                 \L({2 \over \pi} \R)^{3/2},
\EE
where $k$ is the Boltzmann constant;
$e$ and $m_e$ are the electron charge and its mass, respectively; and
$Z$ is the mean ion charge. 
According to Spitzer and H\"arm (1953) (see also Borkowski et al. 1989), 
we calculate the latter from the formula 
\BE
Z=\L.\sum_{i,j}n_{i,j}Z_{i,j}^2\R/ \sum_{i,j}n_{i,j}Z_{i,j},
\EE
where the sums are taken on all ions of all chemical elements.
The Coulomb logarithm is taken according to Cohen et al. (1950)
\BE
\ln\Lambda=\ln \L\{ {3\over 2e^3Z} \,
{(kT)^{3/2}\over \L[\pi n_e(1+Z)\R]^{1/2}}\R\}.
\EE
The function $\xi(Z)$ in ~\req{condcoef} is of the order of unity.
We took the expression for $\xi(Z)$ from the paper by Borkovsky  et al. (1989) (see also Max et al. 1980):
\BE
\label{condappr} 
\xi(Z)=0.95\:{Z+0.24\over 1+0.24Z}\,. 
\EE
The values of $\xi(Z)$, obtained with the aid of this formula are
in good agreement with those found by Spitzer and H\"arm (1953).

Thermal conductivity reaches its saturated value
of $F_{\rm sat}$ even at small temperature gradients, because
the electron mean free path relative to collisions with
ions is large (Imshennik and Bobrova 1997):
\BE
\lambda \sim (kT_e)^2/(Z e^4 n_e \ln\Lambda )
\sim 10^4 T_e^2/ Z n_e \; \mbox{[cm]} \; .
\label{epath} 
\EE
Here $Z$ is the mean charge, defined above.
The mean free path $\lambda$ accounts for an appreciable fraction
of the SNR radius at $T_e \sim 10^7 \; \mbox{K} \sim 1 $ keV 
and can formally even exceed it at higher temperatures.
In section \ref{xray}, we assess the role of magnetic fields
that effectively reduce the electron mean free path.

Note that the ion temperature $T_i$  in the two-temperature
plasma can be so high that ions contribute appreciably to the thermal conductivity (Bobrova and Sasorov 1993). 
We have not yet taken this
effect into account (as in other astrophysical work on
SNRs), but its consideration deserves the attention of
theorists.

\subsection{Radiative losses} \lvm

The importance of radiative losses in metal-rich
supernova ejecta  was emphasized already by Hamilton and Sarazin (1984b).  
These authors have shown that thermal instability and a catastrophic cooling of ejecta could develop already at early stages of an SNR evolution,
for example, with the ejecta parameters, corresponding to the Tycho SNR.
Nevertheless, in subsequent works on  modeling the X-ray emission of the Tycho SNR (Itoh et al. 1988; Brinkmann et al. 1989; Badenes et al. 2003), the
radiative losses were disregarded.

Our calculations do take into account the
radiative losses both in the interstellar medium and
in the ejecta and confirm the conclusions reached
by Hamilton and Sarazin (1984b) that these losses
cannot be ignored without a special test, even for
young SNRs. Figure~\ref{losses} shows, for the W7 and MR0
models, the evolution of the thermal energy and the
rate of its change in two Lagrangian zones, one belonging
to the ejecta and the other -- to the
interstellar medium. The zones were chosen so that
they were simultaneously heated by the reverse and
forward shocks and a significant amount of iron was
contained in the chosen ejecta zone. The models
were computed up to the age of the Tycho SNR,
which is about 430 yr. We see from Fig.~\ref{losses} that the
characteristic cooling time scale $E/ \dot E$ for a gas
element in the interstellar medium is very large $\sim 10^7$ years.

\FIGG losses egasw7ckm15_73_119r egasmrckm15_97_106r [ width=0.5 \textwidth ] 
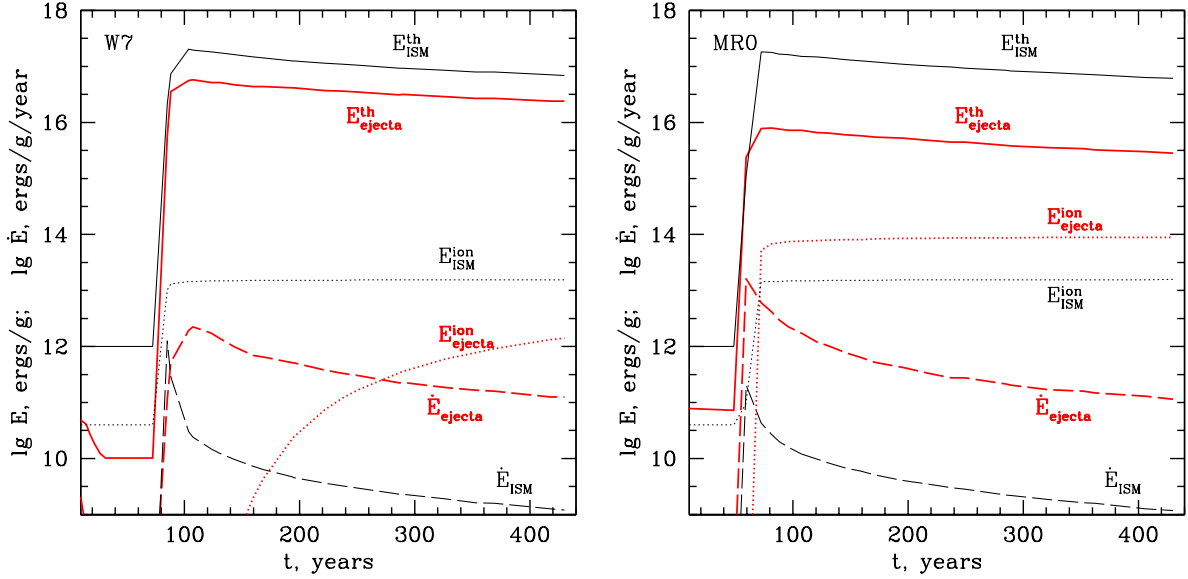
Evolution of the thermal energy (solid lines), the
energy spent on ionization (dotted lines), and the energy
radiated in a year (dashed line) in a mesh
zone of the ejecta versus another one in the ambient medium for the (a)
W7 and (b) MR0 models.

In the ejecta, the thermal energy is an order
of magnitude lower, while the losses are two orders
of magnitude higher; i.e., they may still be considered
small for the two models under consideration.
However, it should be remembered that we have
to average the radial density distribution for our one-dimensional
code. In reality, the supernova ejecta
can be essentially three-dimensional, so there are
dense clumps in the rarefied matter whose cooling
rate is appreciably higher than that of the matter on
average. Such clumps can lose energy on the order of
the thermal energy even by the age of the Tycho SNR.

\FIG instab egasmreqckm15rhop2_97_106r [ width=0.5 \textwidth ] 
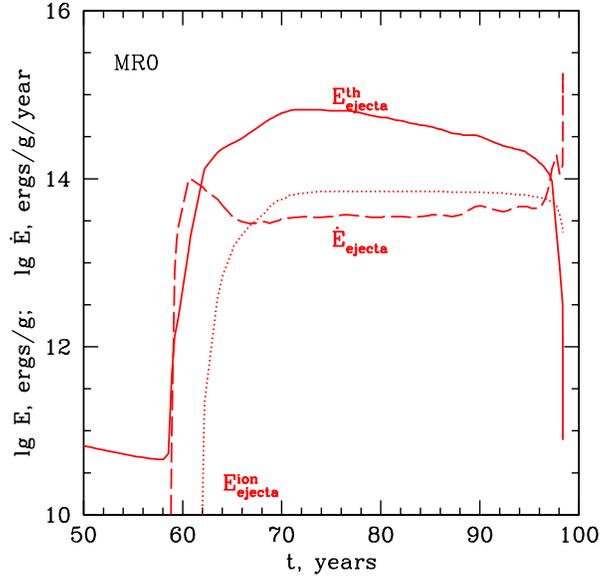
Same as Fig.~\protect\ref{losses} for the MR0 model, but the density
in the displayed zone was increased by a factor of 10,
which leads to the growth of thermal instability in this
zone in less than 100 yr. Here, the model with $T_e=T_i$ is used.

We can illustrate such cooling by artificially increasing the density in one of the outer zones in our models.
The outer layers of the MR0 model are rich in iron, which increases the cooling efficiency.
A threefold increase of the density in one of these zones leads to faster
cooling, than in the original model; still, the cooling time is $\ga 1000$ yr. 
A tenfold increase in density leads to the growth of thermal instability
in this zone in less than 100 yr. This is clearly seen
from Fig.~\ref{instab}, which shows the thermal evolution of the
zone with increased density in the MR0 model with  $T_e=T_i$.
Our calculations also indicate that, in the
model with $T_e \neq T_i$ the extra heating of the electrons
through heat exchange with the ions that are heated
much more strongly at the shock does not give stabilization:
instability grows at the same SNR age. In
contrast, thermal conduction can inhibit the growth
of thermal instability in the electron gas.

Since the outer layers of the W7 model consist of
a carbon-oxygen mixture, they cool more slowly. 
In addition, the expansion velocity in W7 is higher than that in MR0. 
Hence, the density of W7 drops faster, which also increases the cooling time. 
For these reasons, we failed to obtain thermal instability in this model in
1000 yr, while we find the catastrophic cooling in some of well-mixed
models in a few hundred years without any artificial enhancement
of density (if the thermal conduction is suppressed).

Thus, in the one-dimensional models we considered, the cooling proved to be not
very significant for a Tycho-age SNR. 
However, for models with a flatter density
distribution, a lower kinetic energy, or in the three-dimensional case, the
cooling in an ejecta with clumps of gas in the outer layers can be significant,
and the losses cannot be considered negligible on a time scale of several hundred
years.

\subsection{Energy exchange between electrons and ions} \lvm
\label{tie} 

During the passage of a shock front, ions are heated
more strongly than electrons, and their temperature  $T_i$
breaks away from  $T_e$. 
The front structure with thermal conduction when $T_i \ne T_e$ was studied in
detail by Imshennik (1962) and Imshennik and Bobrova
(1997). For purely collisional (Coulomb) thermal
energy exchange, the temperature equilibration
rate in Eqs.~\req{tpe},~\req{tpi} is given by
$$
\nu_{ie}=\frac{2^{5/2}\pi^{1/2} e^4 n_i n_e Z_i^2\ln\Lambda}{m_i m_e
k_b^{3/2}\L({T_i\,/m_i}+{T_e/m_e}\R)^{3/2}}.
$$ 

The question as to the role of collisionless processes
in the energy exchange between the plasma components remains unsolved. 
As is suggested long ago by McKee (1974), the plasma instabilities at the front
can be so strong that the temperatures can become
equal, and in certain cases one gets $T_e>T_i$ according to the calculations by Cargill and Papadopoulos (1988).  See also Lesch (1990), who discussed an efficient $T_e$ --- $T_i$ equilibration mechanism.
On the other hand, the interpretation of the observations for the young
SNR 1006 by Laming et al. (1996) indicates that the
electron temperature remains well below the ion temperature.
The same conclusion ($T_e < 0.1-0.2 T_{\rm mean}$) is reached by Hwang et al. (2002) for the Tycho SNR, however, they do not rule out the presence of a noticeable nonthermal ``tail'' in the  electron distribution either.

To describe the effects of collisionless energy exchange, we
use the following approach.
We introduce a parameter $q_i$ that specifies the fraction of artificial viscosity $Q$, added to the pressure of ions in the equations \req{accel},~\req{tpi}: 
$P_i=P_i({\rm thermal}) + q_i Q$.
In this case for the electronic component we put
$P_e=P_e({\rm thermal}) + (1-q_i) Q$.
If only the collisional exchange is taken into account, then $q_i=1$ and we get
the standard system of equations with the 
heating of only ions at the front (Majorov 1986).

\section{X-ray emission of SNRs under various physical
assumptions} \lvm
\label{xray}

For each supernova model, we compute the SNR evolution under various assumptions,
because the role of plasma effects, magnetic fields, etc., is not yet completely
clear. 
In our calculations, the main cases correspond to the following
assumptions. 
In all cases, the ionization kinetics is fully taken into account.
The losses through radiation: free-free, i.e., Bremsstrahlung, free-bound,
and bound-bound (in X-ray lines), are taken into account at each time step 
for the set of ions available at this step. 
Thus, the radiation depends not only on the
density and temperature, but also on the preceding evolution of the ion
composition. 
The computed SNR spectra were folded with the EPIC PN/XMM-Newton
detector response matrix using the XSPEC software package (Arnaud 1996).

Such processes as the thermal conduction, the plasma heating in shocks, and the
rate of heat exchange can depend on magnetic fields and plasma instabilities.
Therefore, their influence, for example, on the mean free path, was simulated by
varying the parameter $q_i$  and introducing the factor $C_{\rm kill}$ of  $F_{\rm
cond}$ into Eq.~\req{tpe}.  
The factor  $C_{\rm kill}$ ($0 \le C_{\rm kill} \le 1 $) describes 
the fraction of the thermal flux contributed to the system compared
to the standard case~\req{fcond}, where the flux is limited only by the existence
of a maximum speed (the speed of sound) for the heat carriers with no account of a
possible decrease in the particle mean free path due to magnetic fields and
plasma instabilities.

Despite all the uncertainty of the physical processes, we should take into account
the fact that there must be relationships between them. 
For example, the presence
of a magnetic field reduces sharply the particle mean free path: in place of the
Coulomb mean free path \req{epath}, one should use the Larmor radius  
$r_{\rm L}=mcu/eB$. 
Even for weak fields ($B\sim 10^6$ G), we have $r_{\rm L} \ll \lambda$, 
i.e., $C_{\rm kill}$ should be decreased. 
At the same time, the presence of a magnetic field most likely leads to the growth
of plasma instabilities at the front, implying that the electron heating increases
there, and the parameter $q_i$ must decrease.

\begin{table}[h!] 
\caption{Main cases of calculations for each model (W7 and MR0)}
\label{variants}
\begin{center}
\begin{tabular}{|c|c|c|}
\hline
Case & $C_{\rm kill}$  & $q_i$  \\
\hline
basic & 1  &  1  \\
\hline                                                           
C & $10^{-15}$  &  1  \\
\hline                                                           
q & 1  &  0.5  \\
\hline                                                           
Cq &  $10^{-15}$  &   0.5  \\
\hline
\end{tabular}
\end{center}
\end{table}

We have taken the case without a magnetic field ($C_{\rm kill}=1$,
$q_i=1$) to be the basic one. 
We also performed calculations with a reduced $C_{\rm kill}$ and $q_i=0.5$. 
A detailed list of cases and their parameters are given in Table~\ref{variants}. Case Cq can be interpreted as a full allowance for the possible effect of a
magnetic field: it causes extra collisionless heating of the electrons at the shock front, so that the post-shock electron and ion temperatures become almost equal,
and reduces significantly the electron mean free path by suppressing the thermal conduction.
In case C with a magnetic field, the thermal conduction is also suppressed, but it
does not lead to extra non-Coulomb heating at the front for some reasons. 
In case q, the electron heating at the front can be attributed to plasma
instabilities unrelated to the magnetic field; in this case, the thermal
conduction is fully taken into account according to Eq.~\req{fcond}. 
These four cases may be considered as the limiting ones that embrace the entire variety of possibilities encountered in actual SNRs.

\FIG tmpr4w7 w7_5d24_Tr4b [ width=0.9\textwidth ] 
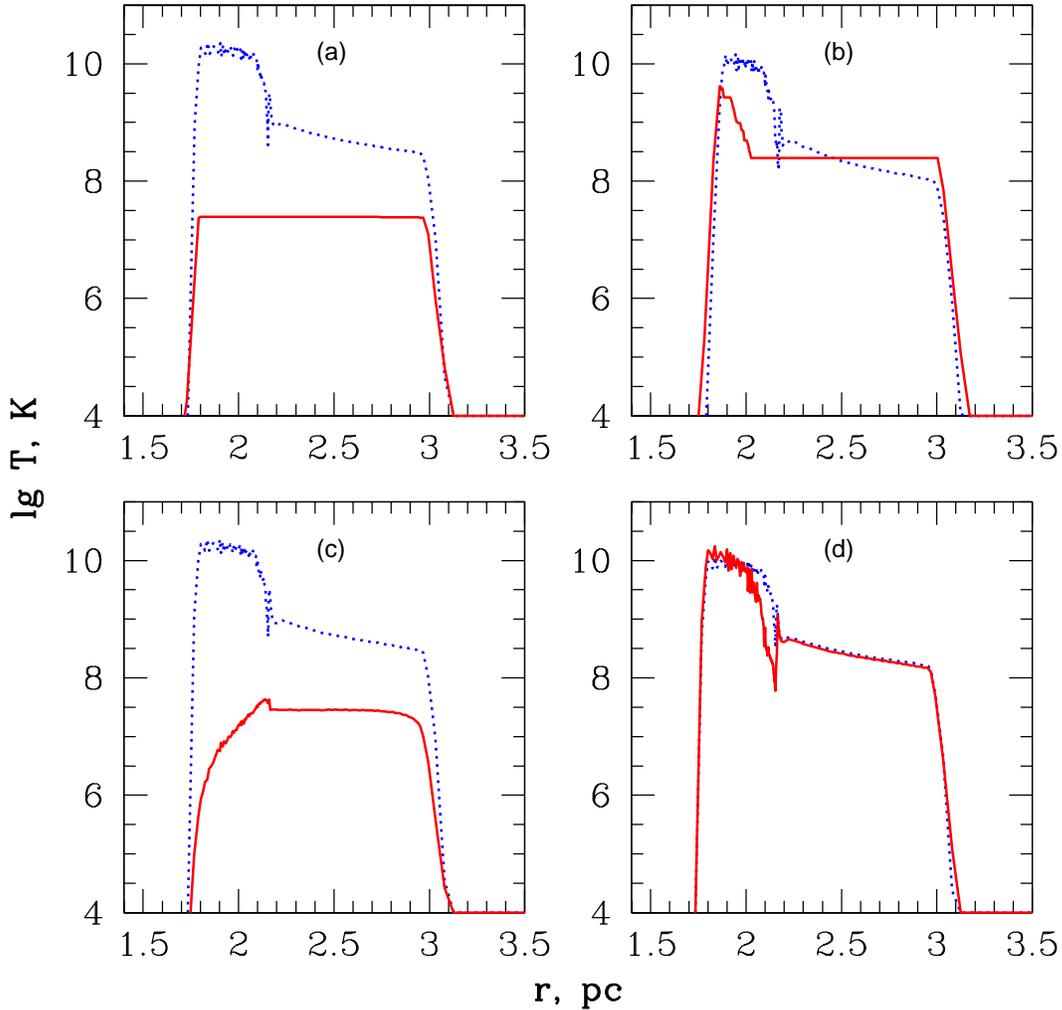
Ion (dotted lines) and electron (solid lines) temperature distributions in the W7 model for four sets of parameters $q_i$ and $C_{\rm kill}$ 
according to Table~\protect\ref{variants}:
(a) basic set, (b) q, (c) C, and (d) Cq.

\FIG tmpr4mr mr_5d24_Tr4b [ width=0.9 \textwidth ] 
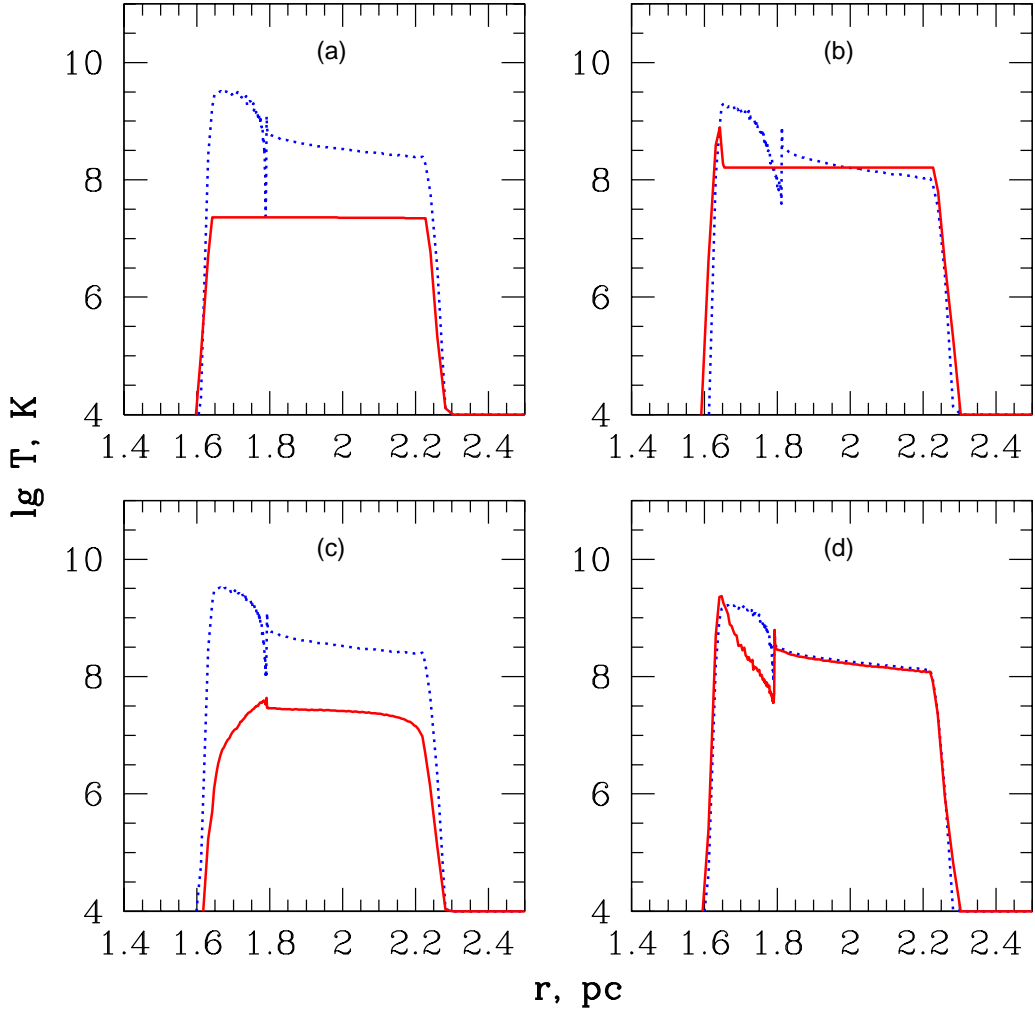
Ion (dotted lines) and electron (solid lines) temperature distributions in the MR0 model for four sets of parameters $q_i$ and $C_{\rm kill}$ 
according to Table~\protect\ref{variants}:
(a) basic set, (b) q, (c) C, and (d) Cq.

Figures~\ref{tmpr4w7} and \ref{tmpr4mr} show the radial distributions of the
electron and ion temperatures for the W7 and MR0 models for an age of 430 yr after
the explosion, which roughly corresponds to the age of the Tycho SNR. 
The behavior of the electron temperature in the ejecta reflects our physical 
models well, and the overall structure of the ejecta is similar for the two
explosion models. 
One can see clearly from the figures how significant the electron thermal conduction can be in young SNRs. 
Thus, in our basic model with cold electrons at the front and
without suppression of the thermal conduction, the electron temperature between
the primary and reverse shocks is completely equalized. 
For strong electron heating at the front, the picture is approximately the same; only a small electron temperature peak is observed in the zone behind the reverse shock. 
Since this spike is not long and the electron temperature rapidly reaches
the value common to the entire SNR, this is unlikely to affect appreciably the ionization rate behind the shock front, but, of course, the equilibrium ionization
state is higher due to the larger dissipation and the higher electron gas temperature at the shock front. 
The case with strong electron heating and suppressed thermal
conduction would have led to the same distribution of the electron and ion
temperatures if radiative losses were not important. 
However, as  can be seen from the figures, the electron temperature in the ejecta
near the contact discontinuity is appreciably lower than the ion temperature. 
The radiative losses are particularly important in the MR0 model with its higher
metallicity, for the outer layers, and density.

The temperature of the shock-heated interstellar medium for any physics differs by
no more than an order of magnitude, while this difference in the ejecta is much
larger; therefore, one might also expect an appreciable effect of the physics on
the radiation. 
Nevertheless, the integrated spectra (Figs.~\ref{spw7} and \ref{spmr}) prove to be
almost insensitive to the physics, except for the continuum level and the small
differences in the ratios of certain spectral lines). 
This is probably because the ionization in a SNR of such an age is time dependent.
Although the the steady-state values of ionization differ markedly for
different temperatures, the ionization rates are similar for all models.
The differences in the spectra being studied can be explained mainly by different chemical composition and radial distributions of chemical elements in different models. 
Since iron lies deeper in W7 than in MR0,
it is ionized later when the ejecta density drop appreciably due to the
expansion (with a higher velocity). 
The ionization rate is much lower because the density is lower. 
As we see from Fig.~\ref{xife},  the highest iron
ion in W7 is FeXXI in the layers where iron is present in
appreciable amounts and ionized most strongly,  while in MR0 it is FeXXV. 
This is the reason why there are significant differences in spectra for different presupernova models.

A clear feature that distinguishes MR0 from W7 is the absence of $K_\alpha$ iron
lines in the W7 spectra in all four cases from Table~\ref{variants} (the same is
true for one-temperature hydrodynamics, $T_e=T_i$).

In our list of lines, the $K_\alpha$ iron line is presented only for 
FeXXIV--FeXXVI. 
Therefore, it comes as no surprise that this line is clearly seen in
the theoretical spectrum (Fig.~\ref{spmr}) in the MR0 model where the ionization
reaches these high ions (see Fig.~\ref{xife}). In the W7 model, the ionization reaches only FeXXI (Fig.~\ref{xife}). 
Nevertheless, it is interesting to assess the role of the excitation
of such iron ions with several $L$-shell electrons: the question remains as to
whether the  $K_\alpha$ line of iron ions will appear as they are excited from the inner $K$-shell.

\FIGFOUR spw7 w7_200_5d24ej99cwac_Xsp w7_200_5d24ej99qi05wacc_Xsp 
w7_200_5d24ej99ckm15cor_Xsp w7_200_5d24ej99qi05ckm15_Xsp 
 [width=0.38 \textwidth, angle=-90 ] 
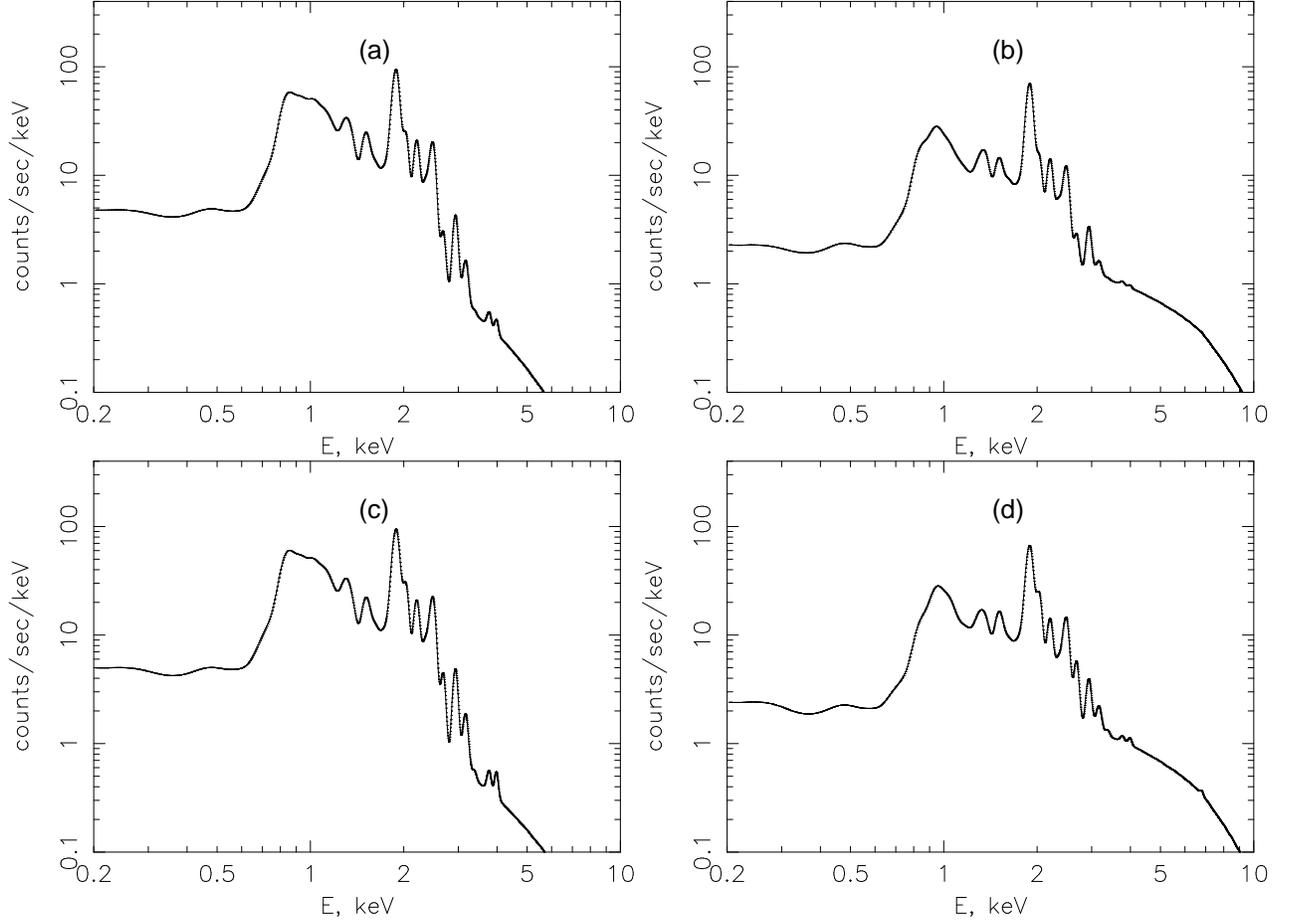
Spectra for the W7 model convolved with the EPIC PN/XMM-Newton detector response matrix: (a) W7$_{\rm basic}$, (b) W7$_{\rm q}$, (c) W7$_{\rm C}$, and 
(d) W7$_{\rm Cq}$.

\FIGFOUR spmr mr_200_5d24te_Xsp mr_200_5d24teqi05wacc_Xsp 
mr_200_5d24ckm15_Xsp mr_200_5d24qi05ckm15_Xsp %
[ width=0.38 \textwidth, angle=-90 ] 
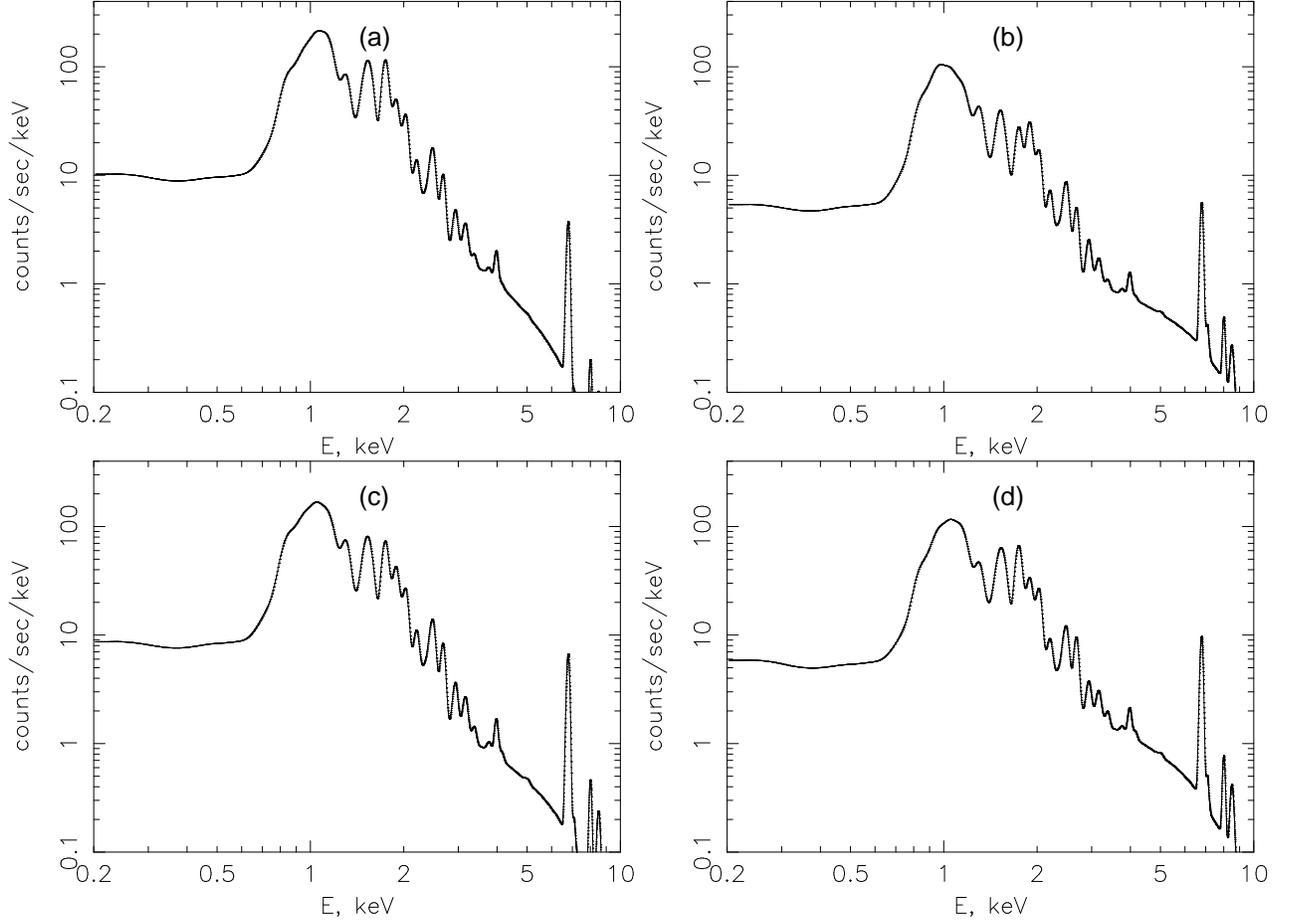
The same as in Fig.~\protect\ref{spw7}, but for MR0 model.

\FIGG xife xife_w7_200_5d24ej99cwac_73 xife_mr_200_5d24te_97 
 [width=0.5 \textwidth ] 
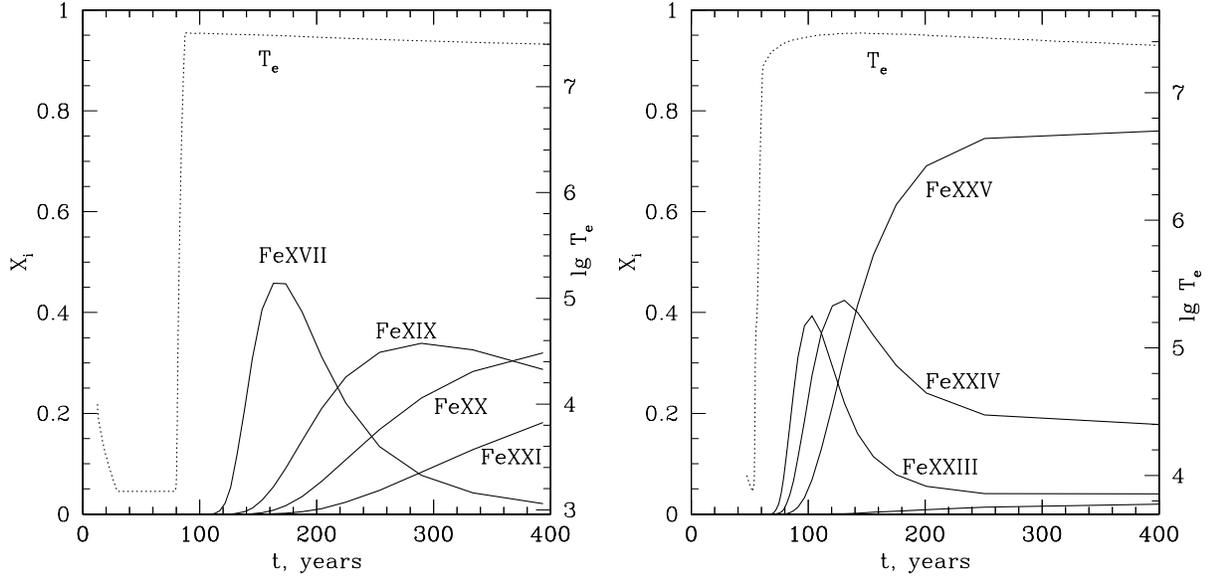
Iron ionization kinetics in the W7 (left) and MR0 (right) models 
in the outermost iron-rich layers of the ejecta (1.02\Msun\
for W7 and 1.37\Msun\ for MR0). 
In the MR0 model, we see the appearance of a hydrogen-like FeXXVI ion. 
The dotted line indicates the evolution of the electron temperature in these zones.

It is clear, that in ``standard'' models with the temperature $T_e$ of the
order $\sim 1$~keV the excitation of innermost shell electrons is
depressed in comparison with the ionization of peripheral electrons by
the exponential factor $\exp(-\Delta E/T_e)$, where 
$\Delta E \sim 7$~keV is the transition energy. 
In the ``hot'' models, the question remains as to whether the excitation or even
the knocking out of $K$-electrons followed by the emission of $K_\alpha$ photons is possible at high mean thermal electron energies or in the presence of a noticeable
nonthermal ``tail'' in their energy distribution. 
Nevertheless, it seems to us that the flux of such photons in W7 is much lower than that in MR0.

It is not easy to find the atomic data on the inner-shell transitions
for iron with several electrons, although such work is conducted both
experimentally and theoretically (see e.g., Goett et al. 1984; Zhang et al. 1990; Ballance et al. 2001; Whiteford et al. 2002).
To place an upper limit  on the impact excitation rate $\langle \sigma v \rangle$
of a transition, we use simple formulas by van Regemorter
(1982) [see e.g. the formula (14.49) from the book by Vainshtein et al., 1979]:
$$
  \langle \sigma v \rangle \sim 6 \times 10^{-8} f_{12} 
\left(\frac{\rm Ry}{\Delta E} \right)^{3/2} \beta^{1/2} \exp(- \beta), 
\quad \beta = -\Delta E/T_e \; ,
$$
where $f_{12}$ is the oscillator strength of the $1 \rightarrow 2$  transition, 
${\rm Ry}=13.6$~eV (in this form this formula is valid for $\beta>1$ 
and agrees well with the more accurate formula (9) from the paper by Sampson and
Zhang (1988). We find from this formula at
$T \sim 10 - 20$~keV, which is typical for the zones of $K_\alpha$ iron lines  formation in ``the hot'' models,
$\langle \sigma v \rangle \sim 10^{-12} \mbox{cm}^3/\mbox{s}$, if we 
take a very high value $f_{12} \sim 0.4$, as in the 
resonance transition of $K_\alpha$ hydrogen-like ion FeXXVI. 
For the inner-shell transitions, for example, in FeXX, the value of $f_{12}$ is significantly lower, and most importantly, 
the electron density $n_e$  in the FeXX excitation region
in W7 is an order of magnitude lower than that
in MR0; i.e., the probability of the excitation of each ion, 
$n_e \langle \sigma v \rangle$, is much lower in W7, which confirms our
conclusion about the low probability of the $K_\alpha$ iron
lines being observable in this model.

The observations by Decourchelle et al. (2001)
indicate that this line is prominent in the Tycho SNR,
but it is weaker than that in the MR0 model. 
The actual ejecta must probably be slightly less mixed
and rapidly expanding than in MR0 or, conversely,
less energetic and more mixed than in W7.

On the other hand, the W7 model better reproduces
the observations from the viewpoint of the
ratio between the Si XIII ($\sim 2$~keV) and Fe XVII
($\sim 0.8$~keV) blends. 
Again, in this case, a slightly
more energetic and less mixed ejecta could also
rectify the situation for MR0.

\FIG 4xprSi 4xprSiK_w7_mr [ width = \textwidth ] 
SiXIII (solid) and FeXVII (dotted) line intensity distributions over the SNR relative to the intensity in the
same lines at the center for the W7 and MR0 models.

\FIG 4xprFe 4xprFeK_w7_mr [ width = \textwidth ] 
Fe~K (solid) and FeXVII (dotted) line intensity distributions over the SNR relative to the intensity in the same lines at the center for the W7 and MR0 models.

Thus, the analysis of the spectra makes it possible to choose an explosion model
that better fits  the observations and, hence, to understand which explosion
mechanism is realized in nature.
Yet it is difficult to determine what physical processes are important inside a
young SNR from the spectrum alone. 
Another opportunity to study the physics of explosion is given to us
by the surface brightness profiles in narrowband
filters corresponding to certain lines of the X-ray
spectrum.
Figure~\ref{4xprSi} demonstrates the surface brightness
distributions in lines of  SiXIII (1.67 -- 2~keV) and Fe XVII
(775 -- 855~eV) for four cases of both
models, while Fig.~\ref{4xprFe} shows the same distributions for the
models in Fe~K (6.2 -- 6.6~keV) and FeXVII lines. 
The distribution in each line is shown relative to its central
intensity. 
Their behavior (except for the Fe K line,
which is almost absent in W7) is similar for both models, when the physical
parameters are changed, although the relative positions of the emitting rings in
different lines do not coincide with the observed one.

\section{Conclusions} \lvm

We have performed for the first time hydrodynamic modeling of the evolution of
young supernova remnants with a self-consistent allowance for the time-dependent
ionization kinetics in a two-temperature plasma and the electron thermal
conduction.
We phenomenologically took into account the influence of magnetic
fields on the thermal conduction and parameterized the effective electron heating at the shock front through plasma processes. 
A large change in thermal conduction
parameters was shown to have a negligible effect on the X-ray spectra that we
obtained by taking into account all of the major elementary processes and by
convolving the theoretically predicted spectra with the EPIC PN/XMM-Newton detector
response matrix.

One of our most important results is that the influence of radiative losses on
hydrodynamics must be taken into account at the earliest evolutionary phases of
SNRs. 
Thus, we have resolved the contradiction that has existed for twenty years
between the results by Hamilton and Sarazin (1984b), who showed the possibility of
thermal instability growing in a metal-rich ejecta, and the complete neglect of
this effect in other works on hydrodynamic modeling of young SNRs. 
In the models we considered, if the heat conduction is suppressed, the thermal instability grows to the point of catastrophic cooling in about 
a thousand years for a threefold density increase in one of the zones, and in less
than 100 years for a tenfold density increase. 
In real three-dimensional models
and when the ejecta interacts with the possible circumstellar matter, thermal
instability can begin to grow much faster without artificial perturbations
(if thermal conduction is not efficient).

Until now, our goal has not been to choose the best model to describe actual
SNRs.  
However, our calculations of the X-ray spectra indicate that supernova
models that lead to the production of iron in the outermost layers such as those
obtained by Dunina- Barkovskaya et al. (2001) and Reinecke et al. (2002) are most
promising for describing the observed iron lines.

The most important effect of those that we disregarded is, of course, the full
three-dimensional calculation of the hydrodynamics. The generation of nonthermal
particles at shock fronts and the evolution and hydrodynamic influence of magnetic
fields should be investigated on a more solid physical basis. 
Nevertheless, it seems to
us that in any future development of the SNR theory, our efficient algorithm of a
self-consistent coupling of hydrodynamics and kinetics of ionization will
retain its significance.

\vskip 1cm

We thank P.V.Sasorov for numerous helpful discussions, 
and V.Zhakhovskii for help in the organizing the calculations.
This work was supported
by the Russian Foundation for Basic Research
(project 02-02-16500a) and by grants from the
Royal Swedish Academy of Sciences at Stockholm
Observatory, Albanova. We are also grateful to
W.Hillebrandt for his support and hospitality at Max Planck Institut
f\"ur Astrophysik (Garching, Germany).

\newpage
 
\centerline{\bf\Large References}
\begin{description}

\item K.A.Arnaud, Astronomical Data Analysis Software and Systems V, ASP 
Conf. Series 101 (Ed. G.Jacoby, J.Barnes, 1996, p.17);\\
{\tt http://xspec.gsfc.nasa.gov/docs/xanadu/xspec/index.html}.

\item M.Arnaud, R.Rothenflug,
   Astron. Astrophys. Suppl. Ser., {\bf  60}, 425 (1985).

\item  O.B.Arushanyan, S.F.Zaletkin, Chislennoe reshenie
 obyknovennykh differentsial'nykh uravnenij na Fortrane.
 Numerical Solution of Ordinary Differential Equations using
 Fortran (MGU, Moscow, 1990) [in Russian].

\item C.Badenes, E.Bravo,  K.J.Borkowski, I.Dom{\'{\i}}nguez,
 Astrophys. J., {\bf 593}, 358 (2003).

\item C.P.Ballance, N.R.Badnell, K.A.Berrington,
J. Phys. B Atomic Mo\-lec\-u\-lar Physics, {\bf 34}, 3287 (2001).

\item S.I.Blinnikov, E.I.Sorokina,
   Astrophys. Space Sci., {\bf 290}, 13 (2004).

\item S.I.Blinnikov, R.Eastman, O.S.Bartunov, V.A.Popolitov,
   S.E.Woos\-ley), Astrophys. J., {\bf 496}, 454 (1998).

\item  N.A.Bobrova, P.V.Sasorov,  Fizika Plazmy, {\bf 19},  789 (1993).

\item K.J.Borkowski, J.M.Shull, C.F.McKee, Astrophys. J., {\bf 336}, 979 (1989).

\item A.V.Borovskij , S.A.Zapryagaev, O.I.Zatserinnyj, and N. L. Manakov,
  Plazma mnogozaryadnykh ionov. Multi-Charged Ion Plasma
 (Khimiya, St.-Petersburg, 1995) [in Russian].

\item  W.Brinkmann, H.~H.Fink,  A.Smith, F.Haberl,
 Astron. Astrophys., {\bf 221}, 385 (1989).

\item P.J.Cargill,  K.Papadopoulos, Astrophys. J., {\bf 329}, L29 (1988).

\item R.A.Chevalier, Astrophys. J. {\bf 258}, 790 (1982). 

\item R.S.Cohen, L.Spitzer~Jr., P.McR.Routly, Phys. Rev., {\bf 80}, 230 (1950).

\item A.Decourchelle et al., Astron. Astrophys., {\bf 365}, L218 (2001).

\item  N.V.Dunina-Barkovskaya, V.S.Imshennik, S.I.Blinnikov, 
Pis'ma v Astron. zhurn., {\bf 27},  412 (2001).

\item C.W.Gear, Numerical Initial Value Problems in Ordinary
      Differential Equations (Englewood Cliffs: Prentice-Hall, 1971).

\item S.J.Goett, D.H.Sampson, R.E.H.Clark,
Astrophys. J. Suppl. Ser., {\bf 54}, 115 (1984).

\item A.J.S.Hamilton, C.L.Sarazin, Astrophys. J., {\bf 281}, 682 (1984a).

\item A.J.S.Hamilton, C.L.Sarazin, Astrophys. J., {\bf 287}, 282 (1984b).

\item U.Hwang, A.Decourchelle, S.S.Holt, R.Petre,
Astrophys. J.  {\bf 581}, 1101 (2002).

\item V.S.Imshennik, Zhurnal vychisl. matem. i matem. fiziki, {\bf 2}, 206 
   (1962).

\item V.S.Imshennik,  N.A.Bobrova, Dinamika stolknovitel'noj
plazmy. Dynamics of Collisional Plasma (M.: Energoatomizdat, 1997)
 [in Russian].

\item H.Itoh, K.Masai, K.Nomoto, Astrophys. J., {\bf 334},  279 (1988).

\item J.M.Laming, J.C.Raymond, B.M.McLaughlin, W.P.Blair,
   Astrophys. J., {\bf 472}, 267 (1996).

\item H.Lesch,  Astron. Astrophys. {\bf 239}, 437 (1990).

\item Majorov S.A.,  Zhurnal vychisl. matem. i matem. fiziki, {\bf 26}, 1735
 (1986).

\item C.E.Max, C.F.McKee, W.C.Mead, Phys. Fluids, {\bf 23}, 1620
   (1980).

\item C.F.McKee, Astrophys. J., {\bf 188}, 335 (1974).

\item  D.K.Nadyozhin, Preprint ITEP-1 (1981).

\item D.K.Nadyozhin, Astrophys. Space Sci., {\bf 112}, 225 (1985).

\item K.Nomoto, F.--K.Thielemann, K.Yokoi, Astrophys. J., {\bf 286}, 644 (1984).

\item H.Nussbaumer, J.Storey,  Astron. Astrophys., {\bf 126}, 75 (1983). 

\item M.Reinecke, W.Hillebrandt, J.C.Niemeyer,  Astron. Astrophys. {\bf 386}, 936 (2002).

\item D.H.Sampson, H.L.Zhang,  Astrophys. J., {\bf  335}, 516 (1988).

\item M.J.Seaton, MNRAS, {\bf 119}, 81 (1959).

\item M.J.Shull, M.Van Steenberg, Astrophys. J. Suppl. Ser., {\bf 48}, 95 (1982).

\item L.Spitzer, R.H\"arm, Phys. Rev. {\bf 89}, 977 (1953).

\item L.A.Vainshtein,  I.I.Sobelman,  E.A.Yukov, Vozbuzhdenie atomov
i ushirenie spektral'nykh linij  (M.: Nauka, 1979).
  I.I.Sobelman,  L.A.Vainshtein, E.A.Yukov,
Excitation of atoms and broadening of spectral line
(Berlin: Springer, 1995).

\item H.Van Regemorter,   Astrophys. J., {\bf  136}, 906 (1962).

\item D.A.Verner,  D.G.Yakovlev, Astrophys. Space Sci., {\bf 165}, 27 (1990). 

\item D.A.Verner,  G.J.Ferland, Astrophys. J. Suppl. Ser.,
  {\bf  103}, 46 (1996).

\item A.D.Whiteford et al.,
J. Phys. B Atomic Molecular Physics, {\bf  35}, 3729 (2002).

\item H.L.Zhang, D.H.Sampson, R.E.H.Clark, Phys. Rev. A, {\bf 41}, 198 (1990).

\end{description}

\end{document}